\RequirePackage{lineno}
\documentclass[aps,preprint,tightenlines,superscriptaddress,byrevtex]{revtex4}
\usepackage{color}
\usepackage{longtable}
\usepackage{graphicx} 
\usepackage{dcolumn}  
\usepackage{pdflscape}
\usepackage{lineno}

\def\NIMA{{ Nucl. Instrum. Methods Phys. Res., Sect. A }}

\begin{document}

\vspace*{-3\baselineskip}
\resizebox{!}{3cm}{\includegraphics{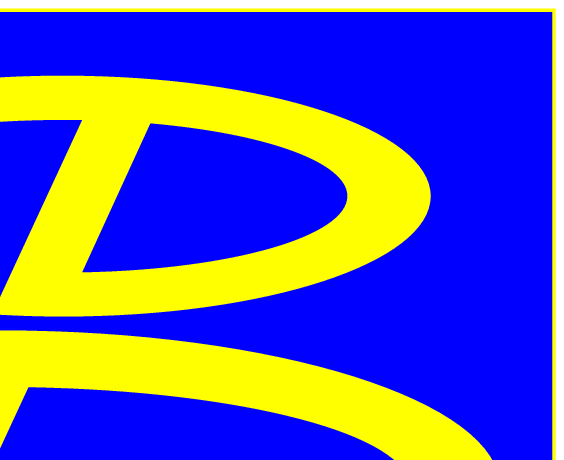}}

\preprint{\vbox{\hbox{Version 0.3}
                \hbox{Belle-CONF-1609}
}}
 
\title{\quad\\[0.5cm] Measurement of $CP$ asymmetry in the $D^{0} \to K^0_S K^0_S$ decay at Belle}

\noaffiliation
\affiliation{Aligarh Muslim University, Aligarh 202002}
\affiliation{University of the Basque Country UPV/EHU, 48080 Bilbao}
\affiliation{Beihang University, Beijing 100191}
\affiliation{University of Bonn, 53115 Bonn}
\affiliation{Budker Institute of Nuclear Physics SB RAS, Novosibirsk 630090}
\affiliation{Faculty of Mathematics and Physics, Charles University, 121 16 Prague}
\affiliation{Chiba University, Chiba 263-8522}
\affiliation{Chonnam National University, Kwangju 660-701}
\affiliation{University of Cincinnati, Cincinnati, Ohio 45221}
\affiliation{Deutsches Elektronen--Synchrotron, 22607 Hamburg}
\affiliation{University of Florida, Gainesville, Florida 32611}
\affiliation{Department of Physics, Fu Jen Catholic University, Taipei 24205}
\affiliation{Justus-Liebig-Universit\"at Gie\ss{}en, 35392 Gie\ss{}en}
\affiliation{Gifu University, Gifu 501-1193}
\affiliation{II. Physikalisches Institut, Georg-August-Universit\"at G\"ottingen, 37073 G\"ottingen}
\affiliation{SOKENDAI (The Graduate University for Advanced Studies), Hayama 240-0193}
\affiliation{Gyeongsang National University, Chinju 660-701}
\affiliation{Hanyang University, Seoul 133-791}
\affiliation{University of Hawaii, Honolulu, Hawaii 96822}
\affiliation{High Energy Accelerator Research Organization (KEK), Tsukuba 305-0801}
\affiliation{J-PARC Branch, KEK Theory Center, High Energy Accelerator Research Organization (KEK), Tsukuba 305-0801}
\affiliation{Hiroshima Institute of Technology, Hiroshima 731-5193}
\affiliation{IKERBASQUE, Basque Foundation for Science, 48013 Bilbao}
\affiliation{University of Illinois at Urbana-Champaign, Urbana, Illinois 61801}
\affiliation{Indian Institute of Science Education and Research Mohali, SAS Nagar, 140306}
\affiliation{Indian Institute of Technology Bhubaneswar, Satya Nagar 751007}
\affiliation{Indian Institute of Technology Guwahati, Assam 781039}
\affiliation{Indian Institute of Technology Madras, Chennai 600036}
\affiliation{Indiana University, Bloomington, Indiana 47408}
\affiliation{Institute of High Energy Physics, Chinese Academy of Sciences, Beijing 100049}
\affiliation{Institute of High Energy Physics, Vienna 1050}
\affiliation{Institute for High Energy Physics, Protvino 142281}
\affiliation{Institute of Mathematical Sciences, Chennai 600113}
\affiliation{INFN - Sezione di Torino, 10125 Torino}
\affiliation{Advanced Science Research Center, Japan Atomic Energy Agency, Naka 319-1195}
\affiliation{J. Stefan Institute, 1000 Ljubljana}
\affiliation{Kanagawa University, Yokohama 221-8686}
\affiliation{Institut f\"ur Experimentelle Kernphysik, Karlsruher Institut f\"ur Technologie, 76131 Karlsruhe}
\affiliation{Kavli Institute for the Physics and Mathematics of the Universe (WPI), University of Tokyo, Kashiwa 277-8583}
\affiliation{Kennesaw State University, Kennesaw, Georgia 30144}
\affiliation{King Abdulaziz City for Science and Technology, Riyadh 11442}
\affiliation{Department of Physics, Faculty of Science, King Abdulaziz University, Jeddah 21589}
\affiliation{Korea Institute of Science and Technology Information, Daejeon 305-806}
\affiliation{Korea University, Seoul 136-713}
\affiliation{Kyoto University, Kyoto 606-8502}
\affiliation{Kyungpook National University, Daegu 702-701}
\affiliation{\'Ecole Polytechnique F\'ed\'erale de Lausanne (EPFL), Lausanne 1015}
\affiliation{P.N. Lebedev Physical Institute of the Russian Academy of Sciences, Moscow 119991}
\affiliation{Faculty of Mathematics and Physics, University of Ljubljana, 1000 Ljubljana}
\affiliation{Ludwig Maximilians University, 80539 Munich}
\affiliation{Luther College, Decorah, Iowa 52101}
\affiliation{University of Maribor, 2000 Maribor}
\affiliation{Max-Planck-Institut f\"ur Physik, 80805 M\"unchen}
\affiliation{School of Physics, University of Melbourne, Victoria 3010}
\affiliation{Middle East Technical University, 06531 Ankara}
\affiliation{University of Miyazaki, Miyazaki 889-2192}
\affiliation{Moscow Physical Engineering Institute, Moscow 115409}
\affiliation{Moscow Institute of Physics and Technology, Moscow Region 141700}
\affiliation{Graduate School of Science, Nagoya University, Nagoya 464-8602}
\affiliation{Kobayashi-Maskawa Institute, Nagoya University, Nagoya 464-8602}
\affiliation{Nara University of Education, Nara 630-8528}
\affiliation{Nara Women's University, Nara 630-8506}
\affiliation{National Central University, Chung-li 32054}
\affiliation{National United University, Miao Li 36003}
\affiliation{Department of Physics, National Taiwan University, Taipei 10617}
\affiliation{H. Niewodniczanski Institute of Nuclear Physics, Krakow 31-342}
\affiliation{Nippon Dental University, Niigata 951-8580}
\affiliation{Niigata University, Niigata 950-2181}
\affiliation{University of Nova Gorica, 5000 Nova Gorica}
\affiliation{Novosibirsk State University, Novosibirsk 630090}
\affiliation{Osaka City University, Osaka 558-8585}
\affiliation{Osaka University, Osaka 565-0871}
\affiliation{Pacific Northwest National Laboratory, Richland, Washington 99352}
\affiliation{Panjab University, Chandigarh 160014}
\affiliation{Peking University, Beijing 100871}
\affiliation{University of Pittsburgh, Pittsburgh, Pennsylvania 15260}
\affiliation{Punjab Agricultural University, Ludhiana 141004}
\affiliation{Research Center for Electron Photon Science, Tohoku University, Sendai 980-8578}
\affiliation{Research Center for Nuclear Physics, Osaka University, Osaka 567-0047}
\affiliation{Theoretical Research Division, Nishina Center, RIKEN, Saitama 351-0198}
\affiliation{RIKEN BNL Research Center, Upton, New York 11973}
\affiliation{Saga University, Saga 840-8502}
\affiliation{University of Science and Technology of China, Hefei 230026}
\affiliation{Seoul National University, Seoul 151-742}
\affiliation{Shinshu University, Nagano 390-8621}
\affiliation{Showa Pharmaceutical University, Tokyo 194-8543}
\affiliation{Soongsil University, Seoul 156-743}
\affiliation{University of South Carolina, Columbia, South Carolina 29208}
\affiliation{Stefan Meyer Institute for Subatomic Physics, Vienna 1090}
\affiliation{Sungkyunkwan University, Suwon 440-746}
\affiliation{School of Physics, University of Sydney, New South Wales 2006}
\affiliation{Department of Physics, Faculty of Science, University of Tabuk, Tabuk 71451}
\affiliation{Tata Institute of Fundamental Research, Mumbai 400005}
\affiliation{Excellence Cluster Universe, Technische Universit\"at M\"unchen, 85748 Garching}
\affiliation{Department of Physics, Technische Universit\"at M\"unchen, 85748 Garching}
\affiliation{Toho University, Funabashi 274-8510}
\affiliation{Tohoku Gakuin University, Tagajo 985-8537}
\affiliation{Department of Physics, Tohoku University, Sendai 980-8578}
\affiliation{Earthquake Research Institute, University of Tokyo, Tokyo 113-0032}
\affiliation{Department of Physics, University of Tokyo, Tokyo 113-0033}
\affiliation{Tokyo Institute of Technology, Tokyo 152-8550}
\affiliation{Tokyo Metropolitan University, Tokyo 192-0397}
\affiliation{Tokyo University of Agriculture and Technology, Tokyo 184-8588}
\affiliation{University of Torino, 10124 Torino}
\affiliation{Toyama National College of Maritime Technology, Toyama 933-0293}
\affiliation{Utkal University, Bhubaneswar 751004}
\affiliation{Virginia Polytechnic Institute and State University, Blacksburg, Virginia 24061}
\affiliation{Wayne State University, Detroit, Michigan 48202}
\affiliation{Yamagata University, Yamagata 990-8560}
\affiliation{Yonsei University, Seoul 120-749}
  \author{A.~Abdesselam}\affiliation{Department of Physics, Faculty of Science, University of Tabuk, Tabuk 71451} 
  \author{I.~Adachi}\affiliation{High Energy Accelerator Research Organization (KEK), Tsukuba 305-0801}\affiliation{SOKENDAI (The Graduate University for Advanced Studies), Hayama 240-0193} 
  \author{K.~Adamczyk}\affiliation{H. Niewodniczanski Institute of Nuclear Physics, Krakow 31-342} 
  \author{H.~Aihara}\affiliation{Department of Physics, University of Tokyo, Tokyo 113-0033} 
  \author{S.~Al~Said}\affiliation{Department of Physics, Faculty of Science, University of Tabuk, Tabuk 71451}\affiliation{Department of Physics, Faculty of Science, King Abdulaziz University, Jeddah 21589} 
  \author{K.~Arinstein}\affiliation{Budker Institute of Nuclear Physics SB RAS, Novosibirsk 630090}\affiliation{Novosibirsk State University, Novosibirsk 630090} 
  \author{Y.~Arita}\affiliation{Graduate School of Science, Nagoya University, Nagoya 464-8602} 
  \author{D.~M.~Asner}\affiliation{Pacific Northwest National Laboratory, Richland, Washington 99352} 
  \author{T.~Aso}\affiliation{Toyama National College of Maritime Technology, Toyama 933-0293} 
  \author{H.~Atmacan}\affiliation{Middle East Technical University, 06531 Ankara} 
% ...
\collaboration{The Belle Collaboration}

\begin{abstract}
We report a measurement of the time-integrated $CP$ asymmetry in the neutral charm meson decay $D^0 \to K^0_S K^0_S$ using 921~fb$^{-1}$ data collected at the $\Upsilon(4S)$ and $\Upsilon(5S)$ resonances with the Belle detector at the KEKB asymmetric-energy $e^+e^-$ collider. The observed asymmetry is
$$ A_{CP}(D^0 \to K^0_S K^0_S) = (-0.02 \pm 1.53 \pm 0.17) \%, $$
where the first uncertainty is statistical and the second systematic. This latter uncertainty is dominated by the error of the normalisation channel. The result is consistent with Standard Model expectations and improves the uncertainty with respect to previous measurement of this quantity by more than a factor of three.
\end{abstract}

\maketitle

Charge-parity violation (CPV) in charmed meson decays has not yet been observed and is predicted to be small [$\mathcal{O}(10^{-3})$] in the Standard Model (SM). Hence, evidence of CPV in charm decays reported by LHCb~\cite{lhcb1} in $D^{0}\rightarrow h^{-}h^{+}$ decays, where $h = K, \pi$, took many by surprise and generated a renewed interest in this field as an observation of large CPV in charm decays could hint at New Physics (NP). The difference between the $CP$ asymmetries in $D \to K^+K^-$ and $\pi^+\pi^-$ decays, $\Delta A_{CP}$, was measured to be ($-0.82 \pm 0.21 \pm 0.11$)\%. Recently, LHCb updated their $\Delta A_{CP}$ result~\cite{lhcb2} and the combined $\Delta A_{CP}$ value~\cite{hfag} is consistent with no CPV at 6.5\% CL. Though there is no current evidence of nonzero asymmetry, CPV in charm decays is investigated in other channels. Singly Cabibbo-suppressed (SCS) decays are of special interest as the possibility of interference with NP amplitudes could lead to larger CPV than predicted
by SM. The $D^{0} \to K^0_{S} K^0_{S}$ decay is one such channel~\cite{hiller}. The most recent SM-based analysis obtained a 95\% confidence level upper limit of 1.1\% for direct $CP$ violation in this decay~\cite{Nierste:2015zra}. The search for $CP$ asymmetry in $D^{0} \to K^0_{S} K^0_{S}$ has been performed first by the CLEO Collaboration~\cite{Bonvicini:2000qm} using a data sample of 13.7 fb$^{-1}$ of $e^+e^-$ collisions at the $\Upsilon(4S)$ with a measured $CP$ asymmetry of $(-23 \pm 19)\%$. Recently, LHCb measured a time-integrated $CP$ asymmetry in $D^{0}\rightarrow K^0_{S}K^0_{S}$ of ($-$2.9 $\pm$ 5.2 $\pm$ 2.2)\%, where the first uncertainty is statistical and the second systematic~\cite{Aaij:2015fua}. The LHCb result is consistent with no CPV, in agreement with SM expectations. 
The Belle Collaboration has amassed a huge number of $e^+e^-$ collisions at the $\Upsilon(4S)$ and $\Upsilon(5S)$ resonances, and hence, can significantly improve the measurement.


The analysis is based on a data sample that corresponds to an integrated luminosity of 921~fb$^{-1}$ collected with the Belle detector~\cite{abashian} at the KEKB asymmetric-energy $e^+e^-$ collider operating at the $\Upsilon$(4S) resonance, $\Upsilon$(4S) off-resonance, and $\Upsilon$(5S) resonance with integrated luminosities 711.0~fb$^{-1}$, 89.4~fb$^{-1}$, and 121.4~fb$^{-1}$, respectively~\cite{kurokawa}. The Belle detector is a large-solid-angle spectrometer, which includes a silicon vertex detector (SVD), a 50-layer central drift chamber (CDC), an array of aerogel threshold Cherenkov counters (ACC), time-of-flight scintillation counters (TOF), and an electromagnetic calorimeter (ECL) comprised of CsI(Tl) crystals located inside a superconducting solenoid coil that provides a 1.5~T magnetic field. An iron flux return located outside the coil is instrumented to detect $K^{0}_{L}$ mesons and identify muons.


 In this paper, we measure the time-integrated $CP$ asymmetry ($A_{CP}$) of the neutral charm meson decays $D^{0}\rightarrow K_{S}^{0}K_{S}^{0}$. The $D^0$ mesons are required to originate from the decay $D^{*+} \to D^0 \pi^+$ in order to provide a tag of the $D$ flavor as well as to suppress combinatorial background.
Assuming the total decay width to be same for particles and antiparticles, the time-integrated asymmetry is: \begin{equation}
   A_{CP}=\frac{\Gamma (D^{0}\rightarrow K_{S}^{0}K_{S}^{0})-\Gamma (\bar D^{0}\rightarrow K_{S}^{0}K_{S}^{0})}{\Gamma (D^{0}\rightarrow K_{S}^{0}K_{S}^{0})+\Gamma (\bar D^{0}\rightarrow K_{S}^{0}K_{S}^{0})},
\end{equation}
where $\Gamma$ represents the partial decay width. Here, the $A_{CP}$ term has the following contributions : \\
\begin{equation}
  A_{CP} = A_{CP}^{d} + A_{CP}^{m} + A_{CP}^{i}, 
\end{equation} 
where, $A_{CP}^{d}$ is the direct-CPV contribution or CPV in decay that is decay-mode dependent, $A_{CP}^{m}$ is the CPV in mixing of $D$-$\bar{D}$, and $A_{CP}^{i}$ is the CPV due to  interference between decays with and without mixing.  Although the last term depends on the phase difference between mixing and decay, it generally depends on the decay mode. However, the decay phase in $D$ decays is always close to zero and $A_{CP} $ is almost universal. The asymmetry in the decay width is
\begin{equation}
  A_{\Gamma}=\frac{\tau(D^{0}\rightarrow K_{S}^{0}K_{S}^{0})-\tau(\bar D^{0}\rightarrow K_{S}^{0}K_{S}^{0})}{\tau(D^{0}\rightarrow K_{S}^{0}K_{S}^{0})+\tau(\bar D^{0}\rightarrow K_{S}^{0}K_{S}^{0})} = -(A_{CP}^{m} + A_{CP}^{i})
\end{equation}
The world average for $A_{\Gamma} $~\cite{eq4}, $(-0.056\pm 0.040)\%$, is consistent with zero~\cite{hfag}. In the SM, indirect CPV ($A_{CP}^{m} + A_{CP}^{i}$) is expected to be very small, of the order of $10^{-3}$, and is universal for $CP$ eigenstates. Direct CPV is predicted to be small as well. It is expected to be negligible in Cabibbo-favored modes but, in SCS
modes, it might plausibly be  $\mathcal{O}(10^{-3})$, and could be even larger for the $D^{0}\rightarrow K_{S}^{0}K_{S}^{0}$ mode, as mentioned in Ref.~\cite{Nierste:2015zra}. 

The extracted raw asymmetry
\begin{equation}
  A_{\rm raw} = \frac{N(D^0) - N(\bar{D}^0)}{N(D^0) + N(\bar{D}^0)}
\end{equation}
is given by $A_{\rm raw} = A_{CP} + A_{FB} + A^{\pm{}}_{\epsilon}$. Here, $A_{FB}$ is the forward-backward production asymmetry, and $A^{\pm{}}_{\epsilon}$ is the asymmetry due to different detection efficiencies for positively and negatively charged pions. Both can be eliminated through a relative measurement of $A_{CP}$, if the charged final-state particles are identical. The chosen normalization mode is $D^0 \to K_S^0 \pi^0$.
We correct for a non-vanishing asymmetry originating from the different strong interaction of $K^0$ and ${\bar{K}}^0$ mesons with nucleons of the detector material, $A^{K}_{\epsilon}$, estimated to be $-0.11\%$ in Ref.~\cite{belle_K}, and assign a residual systematic uncertainty of $0.01\%$. This effect is cancelled in $D^{0}\rightarrow K_{S}^{0}K_{S}^{0}$ decay since $D^{0}$ decays to $K^{0} \bar {K^{0}}$.
The $CP$ asymmetry of the signal mode can then be expressed as 
\begin{equation}
A_{CP}(D^0 \to K_S^0K_S^0) = A_{\rm raw}(D^0 \to K_S^0K_S^0) - A_{\rm raw}(D^0 \to K_S^0\pi^0) + A_{CP}(D^0 \to K_S^0\pi^0) + A^{K}_{\epsilon},
\end{equation}
where $A_{CP}(D^0 \to K_S^0\pi^0)$ is the world average of $CP$ asymmetry of the normalization mode: $A_{CP}(D^0 \to K_S^0\pi^0) = (-0.20 \pm 0.17)\%$~\cite{pdg}.

The analysis procedure is developed using Monte Carlo (MC) simulation based on \textsc{EvtGen}~\cite{EvtGen} and \textsc{GEANT3}~\cite{GEANT} and includes final-state radiation (FSR) effects simulated by PHOTOS~\cite{PHOTOS}. The selection criteria are optimized using a figure of merit, defined as $N_{\rm sig}/\sqrt{N_{\rm sig} + N_{\rm bkg}}$, where $N_{\rm sig}$ ($N_{\rm bkg}$) represents the number of signal (background) events in a defined signal region. We use a large signal MC sample, about a few hundred times more in size than expected in data. To estimate $N_{\rm sig}$, we use ${\cal B}(D^0 \to K_S^0 K_S^0) = 1.8 \times 10^{-4}$~\cite{pdg}. The MC sample used to estimate the background corresponds to a luminosity of six times that of data. The background is scaled by the ratio of the number of events in data and MC in the $\Delta M$ sideband, $0.148$~GeV/$c^{2}$ $< \Delta M < 0.160$~GeV/$c^{2}$, where $\Delta M$ is the mass difference between the reconstructed $D^{*}$ and~$D$.

We require a slow pion ($\pi_{s}$) candidate to originate from near the interaction point (IP) by restricting its impact parameters along and perpendicular to the $z$ axis to be less than 3~cm and 1~cm, respectively. The $z$ axis is defined as the direction opposite the $e^+$ beam. We require that the ratio of the particle identification (PID) likelihood, ${\cal L}_{\pi}/({\cal L}_{\pi} + {\cal L}_{K})$, be greater than 0.4. Here, ${\cal L}_{\pi}$ (${\cal L}_{K}$) is the likelihood of a track being a pion (kaon) and is calculated using specific ionization from the CDC, time-of-flight information from the TOF and the number of photoelectrons in the ACC. With the above PID requirement, the pion identification efficiency is above $95\%$ with a kaon misidentification probability below $5\%$.

$K_{S}^{0}$ mesons are reconstructed from pairs of oppositely charged tracks, treated as pions, using an algorithm based on neural network (NN)~\cite{ksref}. The NN uses the following information: the $K^0_S$ momentum in the laboratory frame; the distance along the $z$ axis between the two track helices at their closest approach; the flight length in the $x-y$ plane; the angle between the $K_S^0$ momentum and the vector joining the $K_S^0$ decay vertex to the IP; the angle between the pion momentum and the laboratory-frame direction in the $K_S^0$ rest frame; the distance-of-closest-approach in the $x-y$ plane between the IP and each pion helix; and the pion hit information in the SVD and CDC. We also require that the reconstructed invariant mass be within 15~MeV/$c^{2}$ (about four times the resolution) of the nominal $K_{S}^{0}$ mass~\cite{pdg}.
We reconstruct neutral pion candidates from pairs of electromagnetic showers in the ECL that are not matched to any charged track. Showers in the barrel (end-cap) region of the ECL must exceed 60 (100)~MeV to be considered as a $\pi^0$-daughter candidate. The invariant mass of the $\pi^0$ candidate must lie within 25 MeV/$c^2$ (about four times the resolution) of the known $\pi^0$ mass~\cite{pdg}. The $\pi^0$ momentum is required to be greater than 640 MeV/$c$.

To reconstruct the $D^{0}$ candidates, we combine two reconstructed $K_{S}^{0}$ (one reconstructed $K^0_S$ and one $\pi^0$ for the normalization mode) and retain the candidates having an invariant mass in the range $1.847$~GeV/$c^2$ $< M < 1.882$~GeV/$c^2$ ($1.758$~GeV/$c^2$ $ < M < 1.930$~GeV/$c^2$). Finally, $\pi_{s}$ is combined with the $D^{0}$ to form a $D^{*}$ candidate, with a requirement on the resulting $\Delta M$ to lie in the range [0.14, 0.16]~GeV/$c^{2}$. We require that the $D^{*+}$ candidate have a momentum greater than 2.2~GeV/$c$ in the centre-of-mass frame. This requirement significantly reduces the combinatorial background.

After applying all the selection criteria, the fraction of events with multiple $D^{*}$ candidates is 8.6$\%$. The $D^{0}$ candidate is selected from the list of candidates as the one having the smallest $\sum \chi^{2}_{K_{S}^{0}}$ , where $\chi^{2}_{K_{S}^{0}}$ is associated with the $K_{S}^{0}$ vertex-constraint fit. In case this $D^{0}$ candidate is common to more than one $D^{*}$ candidate, the candidate having the charged slow pion with the smallest transverse impact parameter is selected. The best candidate selection is found to select correctly the true candidate with an efficiency of  $98\%$. 

The MC-simulated events are used to investigate the sources of background for $D^{0} \to K^0_{S} K^0_{S}$ decays. The physics backgrounds, which are peaking in the $\Delta M$ distribution as the signal, are identified as specific physics process with same final state, $D^{0} \to K_{S}^{0} \pi^+ \pi^-$. Those are estimated directly from data using the $K_{S}^{0}$ mass sideband, $  0.470$~GeV/$c^2$ $ < M_{\pi\pi} < 0.478 $~GeV/$c^2$ and  $ 0.516$~GeV/$c^2$ $ < M_{\pi\pi} < 0.526 $~GeV/$c^2$. 

\begin{figure}
\begin{center}
\includegraphics[width=13cm]{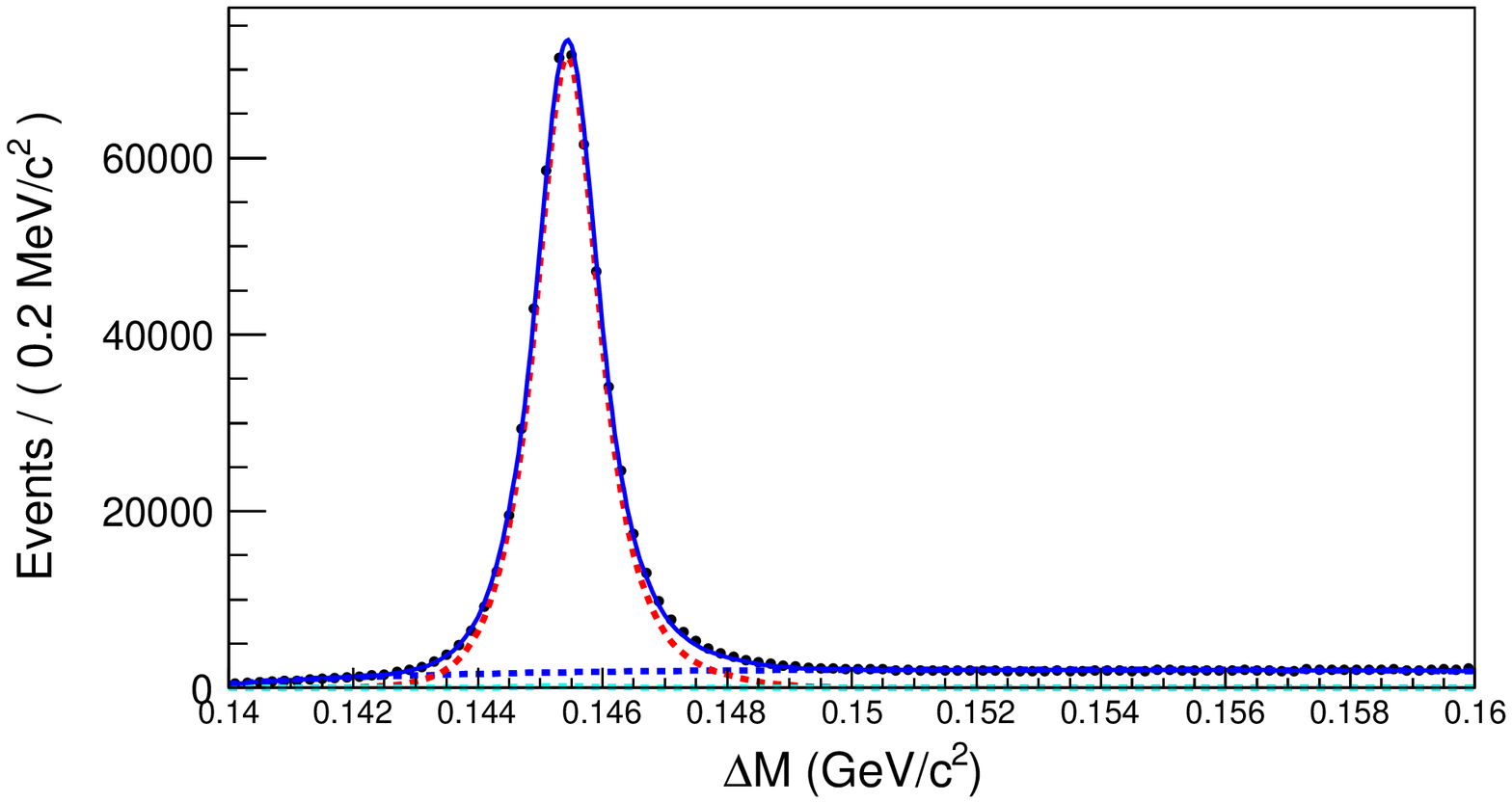} 
\includegraphics[width=13cm]{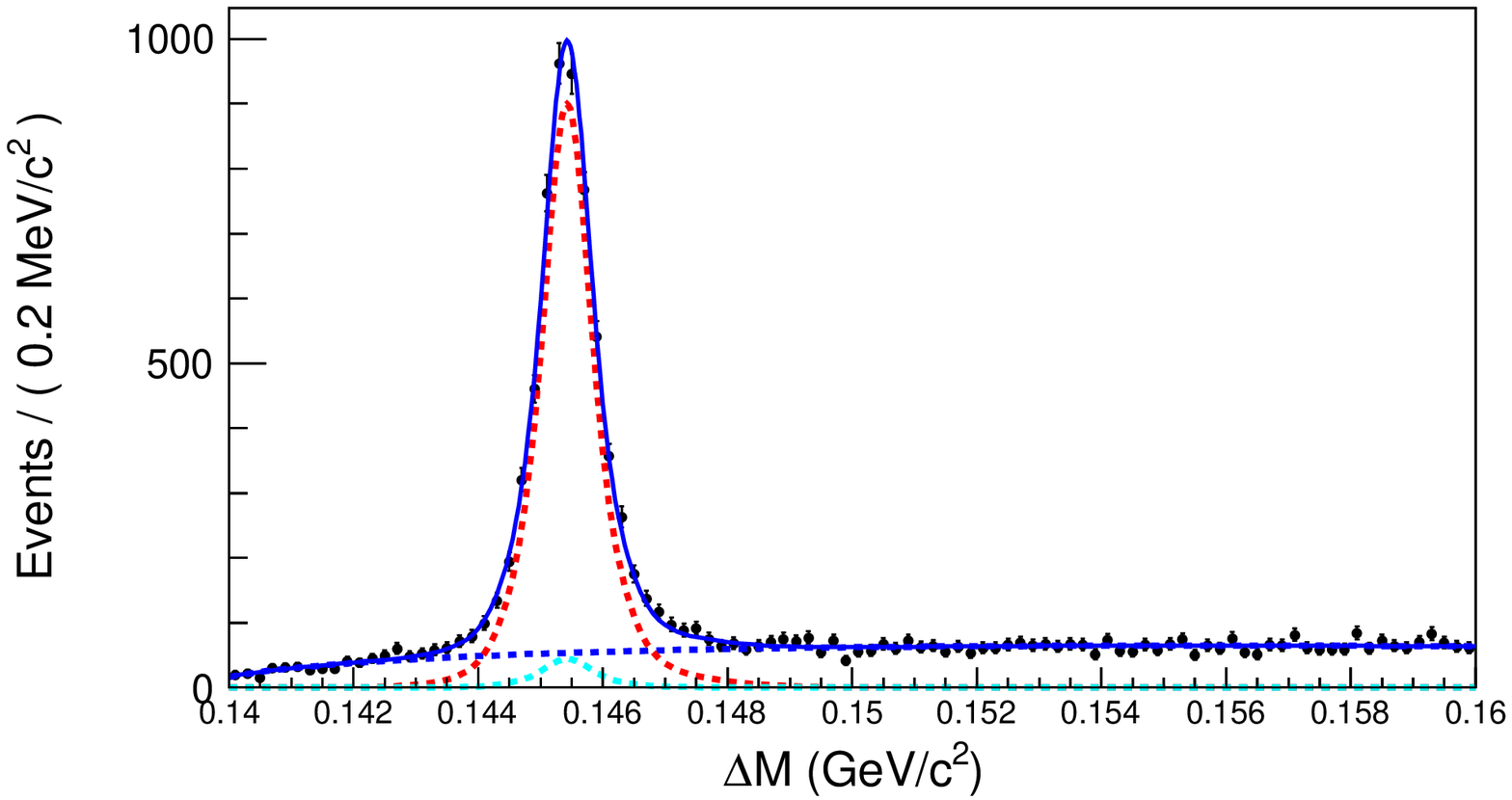}
\caption{Distributions of the mass difference $\Delta M$ for the $K_S^0\pi^0$ (top) and $K_S^0 K_S^0$ (bottom) final states. Points with error bars are the data, the solid curves show the results of the fit, the dashed (blue) curves are the non-peaking background predictions and the dashed (cyan) curves are the peaking background.}
\label{fig:total}
\end{center}
\end{figure}

The $\Delta M $ distributions for $D^0 \to K_S^0 \pi^0$ and $D^0 \to K_S^0 K_S^0$ are shown in Fig.~\ref{fig:total}. We describe the signal shapes by the sum of two symmetric and one asymmetric Gaussian functions with a common mean. Most of the shape parameters are fixed from MC, except for the mean and a width-calibration factor reflecting the possible MC-data difference. The peaking-background component has the same shape as the signal and its yield is fixed to the estimation described above.
The background shapes are modeled with a threshold function: $f(x)$ = $(x-m_{\pi})^{a}\exp[-b(x-m_{\pi})]$, where $m_{\pi}$ is the nominal charged pion mass and $a$ and $b$ are shape parameters. The signal yield for $D^0 \to K_S^0 K_S^0$ is $5399 \pm 87$ events and for $D^0 \to K_S^0 \pi^0$ is $531807 \pm 796$ events.

\begin{figure}
\begin{center}
\includegraphics[width=13cm]{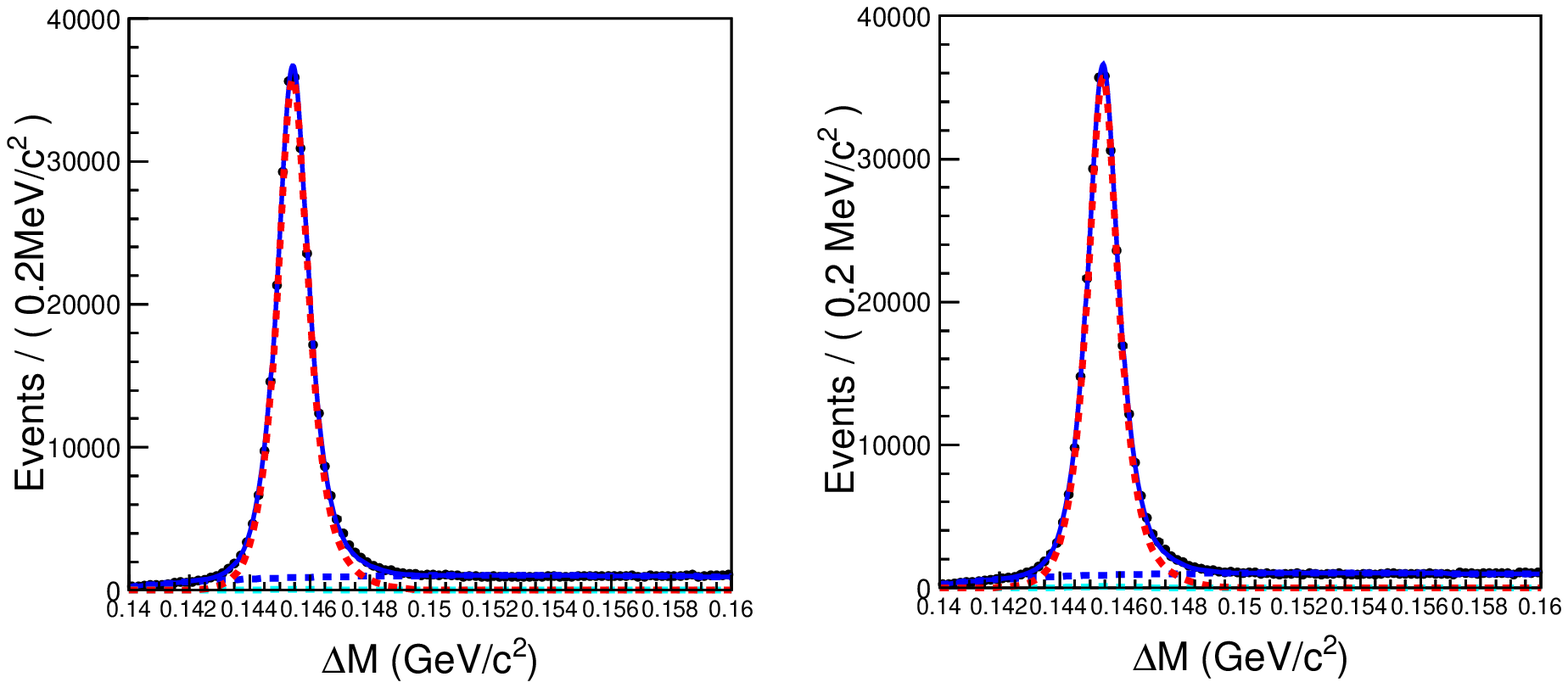}
\includegraphics[width=13cm]{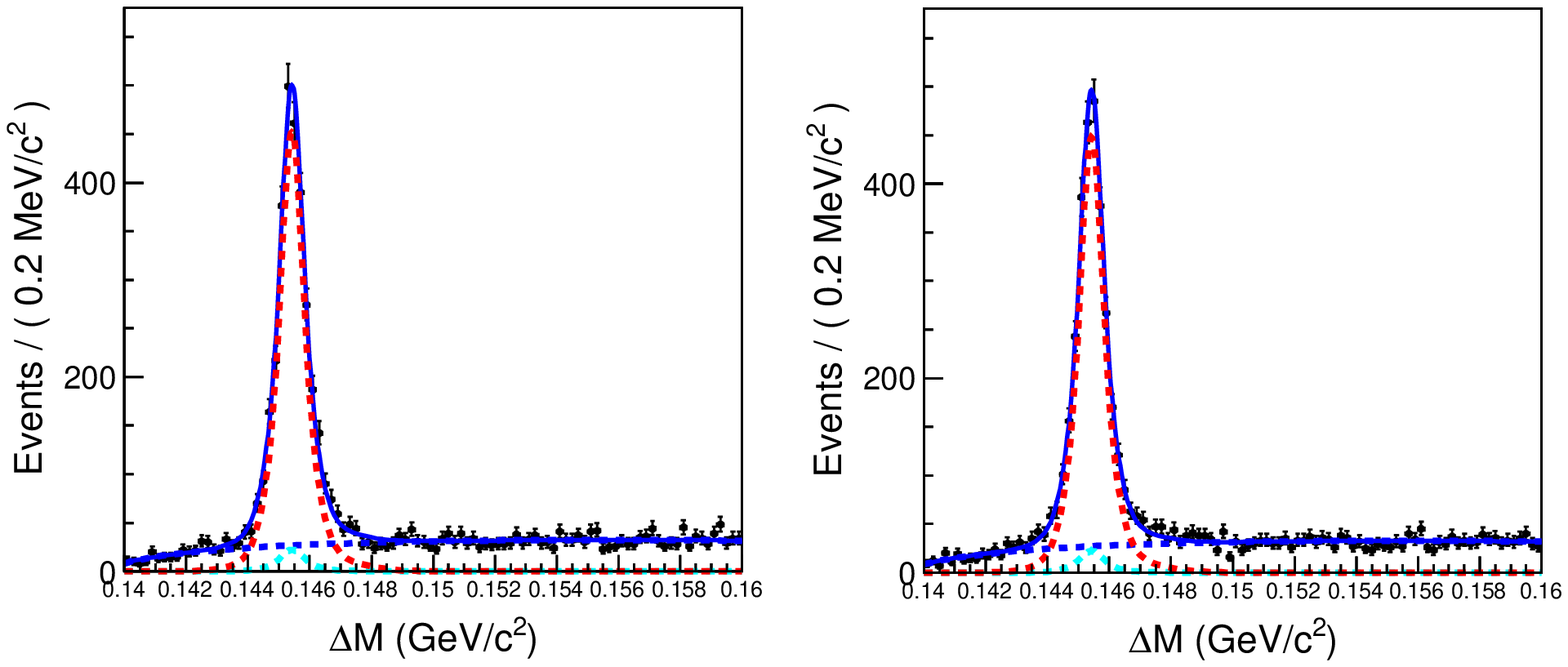}
\caption{Distributions of the mass difference $\Delta M$ for the $K_S^0\pi^0$ (top) and $K_S^0 K_S^0$ (bottom). Left (right) plots are for the $D^{*+}$ ($D^{*-}$) sample. Points with error bars are the data; the curves are explained in the caption of Fig.~\ref{fig:total}. }
\label{fig:markern}
\end{center}
\end{figure}

 A simultaneous fit of the $\Delta M$ for $D^{*+}$ and $D^{*-}$ is used (Fig.~\ref{fig:markern}) to estimate the asymmetry. The signal and background shape parameters are common for both the particle and antiparticle. Both asymmetries in signal and background are allowed to vary in the fit. The $A_{\rm raw}$ for the peaking component in $D^{0}\rightarrow K_{S}^{0}K_{S}^{0}$ case (mostly due to $D \to K_S^{0} \pi^+ \pi^-$) is fixed from the value obtained in data for the $D^{0}\rightarrow K_{S}^{0}\pi^{0}$ signal. The effect due to possible $CP$ asymmetry in the $K_{S}^{0}\pi^{+}\pi^{-}$ and $K_{S}^{0}\pi^{0}$ is accounted for as a systematic uncertainty. The $A_{\rm raw}$ observed in data for $D^{0}\rightarrow K_{S}^{0}K_{S}^{0}$ and $D^{0}\rightarrow K_{S}^{0}\pi^{0}$ is $(+0.45 \pm 1.53)\%$ and $(+0.16 \pm 0.14)\%$, respectively. This gives the time-integrated $CP$-violating asymmetry $A_{CP}$, with only statistical error in the $D^{0}\rightarrow K_{S}^{0}K_{S}^{0}$ decay of
$$
A_{CP} = (-0.02 \pm 1.53) \%
$$
using a data sample of 921~fb$^{-1}$ integrated luminosity.
 
The systematic uncertainties are summarized in Table \ref{tab:tablef}. We identify four sources of systematic uncertainty. The first is due to the uncertainties of the signal shapes: a systematic uncertainty is ascribed for each parameter determined and fixed from MC. We also vary the model by including an additional fudge factor that is allowed to vary. The peaking background yield is determined and fixed from the $K_S^0$ mass sideband. The fit procedure is repeated with its yield varied by its statistical uncertainty. 
We correct for a non-vanishing asymmetry originating from the different strong interaction of $K^0$ and ${\bar{K}}^0$ mesons with nucleons of the detector material, estimated to be $-0.11\%$~\cite{belle_K}, and assign an additional systematic uncertainty of $0.01\%$. For the raw asymmetry, it is fixed from the $K_S^0 \pi^0$ measurement. The dominant systematic uncertainty comes from the uncertainty on the $A_{CP}$ measurement of the normalization channel, $K_S^0 \pi^0$.
Finally, we add these individual contributions in quadrature to obtain the total systematic uncertainty for the time-integrated $CP$-violating asymmetry $A_{CP}$ in the $D^{0}\rightarrow K_{S}^{0}K_{S}^{0}$ decay of $\pm 0.17 \%$.

\begin{table}[h]
\begin{center}  
\begin{tabular}{l | c }  \hline\hline
Source &  Systematic uncertainty, in \% \\ \hline
Signal shape & $\pm 0.01$ \\
Peaking background & $\pm 0.01$  \\
$K^{0}/\bar{K^{0}}$ material effects & $\pm 0.01$ \\
$A_{CP}$ measurement of $K_S^0\pi^0$ & $\pm 0.17$ \\
\hline
 Total & $\pm 0.17$ \\\hline\hline
\end{tabular}
\caption{Summary of various sources of systematic uncertainties in $A_{CP}(D^0 \to K^0_SK^0_S)$.}
 \label{tab:tablef}
\end{center}
\end{table}


In summary, we have measured the time-integrated $CP$-violating asymmetry $A_{CP}$ in the $D^{0}\rightarrow K_{S}^{0}K_{S}^{0}$ decay of
$$
A_{CP} = (-0.02 \pm 1.53 \pm 0.17) \%
$$
using a data sample of 921~fb$^{-1}$ integrated luminosity. The dominant systematic uncertainty arises from the $A_{CP}$ error of the normalisation channel. The result is consistent with the Standard Model expectation and is a significant improvement compared to the previous measurements of CLEO~\cite{Bonvicini:2000qm} and LHCb~\cite{Aaij:2015fua}, already probing the region of interest for NP. 

\hspace{20 mm}

We thank the KEKB group for the excellent operation of the
accelerator; the KEK cryogenics group for the efficient
operation of the solenoid; and the KEK computer group,
the National Institute of Informatics, and the 
PNNL/EMSL computing group for valuable computing
and SINET4 network support.  We acknowledge support from
the Ministry of Education, Culture, Sports, Science, and
Technology (MEXT) of Japan, the Japan Society for the 
Promotion of Science (JSPS), and the Tau-Lepton Physics 
Research Center of Nagoya University; 
the Australian Research Council;
Austrian Science Fund under Grant No.~P 22742-N16 and P 26794-N20;
the National Natural Science Foundation of China under Contracts 
No.~10575109, No.~10775142, No.~10875115, No.~11175187, No.~11475187
and No.~11575017;
the Chinese Academy of Science Center for Excellence in Particle Physics; 
the Ministry of Education, Youth and Sports of the Czech
Republic under Contract No.~LG14034;
the Carl Zeiss Foundation, the Deutsche Forschungsgemeinschaft, the
Excellence Cluster Universe, and the VolkswagenStiftung;
the Department of Science and Technology of India; 
the Istituto Nazionale di Fisica Nucleare of Italy; 
the WCU program of the Ministry of Education, National Research Foundation (NRF) 
of Korea Grants No.~2011-0029457,  No.~2012-0008143,  
No.~2012R1A1A2008330, No.~2013R1A1A3007772, No.~2014R1A2A2A01005286, 
No.~2014R1A2A2A01002734, No.~2015R1A2A2A01003280 , No. 2015H1A2A1033649;
the Basic Research Lab program under NRF Grant No.~KRF-2011-0020333,
Center for Korean J-PARC Users, No.~NRF-2013K1A3A7A06056592; 
the Brain Korea 21-Plus program and Radiation Science Research Institute;
the Polish Ministry of Science and Higher Education and 
the National Science Center;
the Ministry of Education and Science of the Russian Federation and
the Russian Foundation for Basic Research;
the Slovenian Research Agency;
Ikerbasque, Basque Foundation for Science and
the Euskal Herriko Unibertsitatea (UPV/EHU) under program UFI 11/55 (Spain);
the Swiss National Science Foundation; 
the Ministry of Education and the Ministry of Science and Technology of Taiwan;
and the U.S.\ Department of Energy and the National Science Foundation.
This work is supported by a Grant-in-Aid from MEXT for 
Science Research in a Priority Area (``New Development of 
Flavor Physics'') and from JSPS for Creative Scientific 
Research (``Evolution of Tau-lepton Physics'').


\end{document}